\newcommand{\ket}[1]{\left|#1\right\rangle}
\newcommand{\bra}[1]{\left\langle#1\right|}
\newcommand{\ketbra}[3]{\left| #1 \right\rangle_{#3}\!\left\langle #2\right|}
\newcommand{\braket}[3]{\left\langle #1 | #2 \right\rangle_{#3}}

\documentclass[aps,prl,reprint,groupedaddress]{revtex4-2}

\usepackage{amsmath}
\usepackage{mathtools}
\usepackage{xcolor}
\usepackage{graphicx}

\begin{document}


\title{Self-Sifting quantum key distribution}



\author{Saman Sarshar}
\email[]{saman.sarshar@shahroodut.ac.ir}
\affiliation{Faculty of Physics, Shahrood University of Technology, P.O. Box 3619995161,  Shahrood, Iran}
\author{Mostafa Annabestani}
\email[]{annabestani@shahroodut.ac.ir}
\affiliation{Faculty of Physics, Shahrood University of Technology, P.O. Box 3619995161,  Shahrood, Iran}

\date{\today}

\begin{abstract}
In this paper, we introduce a novel two-way quantum key distribution (QKD) protocol in which the sender (Alice) and receiver (Bob) employ one qubit of a maximally entangled Bell state as the quantum channel for key exchange. The protocol incorporates a new security mechanism based on a scrambling operator. Unlike conventional two-way QKD protocols, all sifting operations and eavesdropper detection procedures are postponed until the completion of the quantum communication stage and are performed exclusively by Bob. Since the control mode is never publicly announced, attacks that rely on mode-dependent adaptations or attempt to remain hidden within the control mode are inherently prevented. Furthermore, the traveling qubit does not directly encode key information, substantially limiting the information that can be extracted from attacks targeting the quantum channel alone. An additional distinctive feature of the protocol is that rounds that would ordinarily be discarded can instead be utilized to detect the presence of an eavesdropper. We analyze a broad class of ancilla-based attacks, in which an eavesdropper couples an ancillary system to the transmitted qubit in an attempt to gain information about the key, and show that such attacks are detectable in their most general form.
\end{abstract}


\maketitle

\section{Introduction}
Quantum key distribution (QKD) is one of the most mature applications of quantum information science, enabling two distant parties to establish a shared secret key whose security is guaranteed by fundamental principles of quantum mechanics \cite{renner2008,pirandola2020}. Since the pioneering BB84 protocol \cite{bennett1984} and the entanglement-based E91 scheme \cite{ekert1991}, QKD has developed into a broad field with extensive theoretical and experimental progress\cite{Zhang2025,Shao2025,Kanitschar2025,Liu2023}. Modern security analyses incorporate finite-key effects, coherent attacks, and device imperfections, significantly narrowing the gap between theoretical security proofs and practical implementations \cite{fung2010,curty2019,xu2020,lucamarini2018,pirandola2021}.

Among QKD schemes, two-way protocols are attractive because they enable deterministic key generation without basis reconciliation or sifting, potentially improving efficiency and secret-key rates \cite{Bostrom2002,Lucamarini2014,Beaudry2013}. Their performance, however, is limited by channel loss and by tradeoffs in eavesdropper detection \cite{han2014,Wojcik2003}. In existing schemes, security verification is implemented either through control-mode protocols with test rounds announced during transmission \cite{Bostrom2002,Lucamarini2005}, or through prepare-and-measure schemes such as BB84 where transmitted states encode raw key information and verification is performed only after transmission \cite{bennett1984}. This public availability of control-mode information can provide Eve with additional side information, allowing her to condition or optimize her attack strategy based on the announced structure of the protocol, and has been exploited in several analyses of two-way protocols, including man-in-the-middle–type attacks \cite{Pavacic2021}. In the latter case, delaying sifting and verification implies that any information leakage during transmission may only be revealed after the fact, requiring careful security analysis to bound Eve’s accessible information, particularly in direct communication protocols.


Here we introduce a self-sifting two-way QKD protocol that overcomes these limitations. The protocol eliminates round-by-round control announcements and avoids public comparison of raw-key data. No key information is directly exchanged between the communicating parties, and eavesdropping detection is performed entirely through post-processing at the receiver side, without requiring any additional communication to the sender. In addition, data that are typically discarded in conventional QKD protocols are incorporated into the security verification procedure, improving the efficiency of eavesdropper detection. As a result, the legitimate users can identify the presence of an eavesdropper prior to key acceptance, while also avoiding the need for key-sacrificing rounds in the standard sense.Furthermore, the scrambling mechanism incorporated into the protocol ensures that attack strategies relying on impersonation of a legitimate participant—including man-in-the-middle (MITM) attacks—are fundamentally inapplicable, since any such attempt necessarily induces detectable disturbances. This eliminates the message-mode vulnerability \cite{Pavacic2021} exploited in earlier two-way protocols such as Ping-Pong \cite{Bostrom2002} and LM05 \cite{Lucamarini2005}.


Since the transmitted quantum state carries no direct information about the key, information-extraction attacks that rely on reading the key directly from the traveling qubit are fundamentally inapplicable in the proposed protocol. Beyond such direct attacks, a broad and well-studied class of strategies is ancilla-based attacks \cite{Wojcik2003,Lucamarini2014}, in which an eavesdropper couples an auxiliary quantum system to the exchanged state and attempts to gain information while minimizing the probability of detection. We analyze this class of attacks and prove the security of the protocol against a general ancilla-based attack model. As an explicit example, we show that the double-CNOT (DCNOT) attack—known to be effective against several previously proposed two-way QKD protocols \cite{Lucamarini2014}—cannot be successfully applied to the present scheme, and that the protocol remains secure even against its generalized Double-UNOT extension.

\section{Protocol}
In this protocol the receiver (Bob) has a two-qbit maximally entangled state (Bell states)
\begin{eqnarray}\label{BellStates}
	\ket{\beta}_{pf}=\frac{1}{\sqrt{2}}\left(\ket{0f}+(-1)^p\ket{1{f\oplus1}}\right)
\end{eqnarray}
where p (phase bit) and f (family bit) are $\{0,1\}$ and $\oplus$ used for binary mod.
He ask sender (Alice) decodes her secret bit on it. The protocol's steps are:\\

\textbf{sharing stage:}
\begin{itemize}
	\item[p1] Bob randomly prepare a Bell state and write $(p,f)$.
	\item[p2] Bob keeps one qbit of the Bell state on his side (home qbit) and randomly apply  the unitary scramble operator ($S$) on the other and send it to Alice (travel qbit).
	\item[p3] Alice randomly apply  the unscramble operator ($S^{\dagger}$) on the received qbit and then apply the coding operator $Z^a$ for code $a=\{0,1\}$ and send it back to Bob.
	\item[p4] Bob perform the Bell measurement on both travel and home qbits and write the outcome bits ($p^\prime, f^\prime$). 
	\item[p5] They repeat p1 to p4, $N\geq 2n$ times, for the secret key of length $n$.
\end{itemize}
after this stage, Bob start the\\
\textbf{sifting stage :}
\begin{itemize}
	\item[s1] Bob asks Alice to announce whether she applied ($S^{\dagger}$). He then classifies each round as follows: rounds in which both parties apply the scrambling operators (S and $S^{\dagger}$) or neither party applies them are designated as "code" rounds, corresponding to the Double-Scramble (DS) and Non-Scramble (NS) cases, respectively. Rounds in which only one party applies a scrambling operation are designated as "garbage" rounds (Mono-Scramble (MS)). If Bob intends to employ s4 and s5, he requests Alice to disclose her encoded bit ($a$) for each garbage round; however, to maintain the privacy of all modes, this disclosure may be postponed until immediately before s4.
	 
	\item[s2] Bob check all "code" results and calculate the probability ($Pd_c=\frac{\left\|C\right\|}{N_c}$), where 
	\begin{equation}
	C=\left\{\left(p^\prime,f^\prime\right)| f^\prime \ne f  \right\}
	\end{equation}
	 , $\left\|C\right\|$ is number of $C$ elements and $N_c$ is number of code.
	\item[s3] If $Pd_c \ne 0$, Eve is on the road and they terminate the protocol. 
	\item [s4] Bob check all garbage results and calculate the probability ($Pd_g=\frac{\left\|G\right\|}{N-N_c}$), where
	\begin{eqnarray}
		G=\{\left(p^\prime,f^\prime\right)|& \,\left( p^\prime\ne p\oplus1\oplus a , f^\prime = f\right) \\\nonumber
		&\vee \left( p^\prime\ne p\oplus a , f^\prime = f \oplus 1\right)\}
	\end{eqnarray}
	\item[s5] If $Pd_g \ne 0$, Eve is on the road and they terminate the protocol.
	\item[s6] The protocol is finished and the remaining bits in code cases are secure.
\end{itemize} 

\section{discussion}
In this protocol we use four orthogonal two-qbit maximally entangled states (Bell state)
\begin{eqnarray}
	\ket{\beta_{pf}}=\frac{1}{\sqrt{2}}\left(\ket{0f}+(-1)^p\ket{1{f\oplus1}}\right)
\end{eqnarray}
 as a tool to encode and decode the secret key. These states have valuable features.

 First, the local state of each single qbit has not any information about the Bell state, because \begin{equation}\label{partial-trace}
 	Tr_1\left(\ket{\beta_{pf}}\bra{\beta_{pf}}\right)=Tr_2\left(\ket{\beta_{pf}}\bra{\beta_{pf}}\right)=I
 \end{equation}  
 in which $Tr_i$ refer to the partial trace over the qbit $i$ and $I$ is identity matrix. And second, local operations on a single qbit of a Bell state, can reproduce all others (with a global phase)
  \begin{equation}\label{local-pauli}
  	 Z^mX^n \otimes I \ket{\beta_{pf}}=\pm\ket{\beta_{p\oplus m f\oplus n}}
  \end{equation}  
  where $Z$ and $X$ are Pauli matrix along $z$ and $x$ direction respectively. In fact $Z$, flip the phase $p$ and never change the family $f$ while $X$, flip the family $f$ and never change the phase $p$
  
In this protocol Bob prepare a random Bell state and keeps the first qbit at home (home qbit) and send the second qbit to Alice (travel qbit), so from Eq. \ref{partial-trace}, not only travel qbit from B to A has not any information about the prepared Bell state, the backward qbit A to B (after applying the code operator by Alice) has not any information about code as well. So any attempt of Eve to gain information from travel qbit by direct measurement, are completely failed. 

In other words, the information encodes on the correlation of the whole system (Bell state) and never exchange between A and B. This advantage robust this protocol against the attacks, which have designed to extract the information from exchanging particle, same as intercept and resend \cite{Gisin2002} or PNS \cite{Brassard2000}.

Thus, Eve’s attacks are effectively restricted to strategies aimed at inferring Alice’s encoding operation without directly measuring the traveling qubit; however, such attacks remain detectable within the proposed protocol. Unlike some existing protocols, the control mode (scramble modes) remains completely private throughout the protocol. Although Alice’s encoded bit ($a$) for MS rounds may be disclosed later when evaluating s4 and s5, this occurs only after Eve has already been tested and potentially detected in s3, rendering the disclosed information useless to her. Consequently, Eve cannot determine whether a given transmission corresponds to the DS, MS, or NS case during the protocol and therefore cannot adapt her strategy accordingly. As a result, she must apply a uniform strategy across all cases, which necessarily leads to detectable disturbances. Moreover, any attempt to infer Alice’s code induces either a detectable disturbance or incomplete information gain, since Alice applies a random unscrambling operation prior to encoding. A detailed proof is provided in the next section.

\section{security}\label{security}   
The critical point for the security of this protocol, is the scramble operator $S$. Without $S$, Eve can use several attacks (DCNOT \cite{Lucamarini2014}, Non-orthogobal measurement \cite{Lucamarini2005}) to detect the code of Alice without or minimum detection. 

The scramble operator $S$, must have some properties to guarantee the security of protocol
\begin{itemize}
	\item Produces Quantum Bit Error Rate (QBER), which represents the incorrect results that $B$ would never expect in the absence of Eve.
	\item Produces Silent Ambiguity (SA), which makes ambiguity for Eve in interpreting the results of her attack. This ambiguity must be silent, meaning that Eve should never see any signs of application $S$ in the information she gathers during her attack.

\end{itemize}
The most general form of $S$ can be a $SU(2)$ operator,
\begin{equation}\label{SU2}
S =	U(\alpha,\beta,\theta)=\left( \begin{array}{cc}
		e^{i \alpha} \cos(\theta) & e^{i \beta} \sin(\theta) \\
		-e^{-i \beta} \sin(\theta) & e^{-i \alpha} \cos(\theta)
	\end{array}\right)
\end{equation}
which can be represented in the Bloch-representation form as 
$S=\sum_{i=0}^{3}{r_i\sigma_i}$ in which $\sigma_i$ are Pauli matrices where $\sigma_0\equiv I$ and $i=1,2,3$ used for $X,Y,Z$ respectively. $r_i=\frac{1}{2}Tr\left(\sigma_iS\right)$ is amplitude of applying Pauli matix $\sigma_i$. So applying $S$ on a Bell state change the family of state with probability
$P_{c}=\left|r_1\right|^2+\left|r_2\right|^2=\sin\left(\theta\right)^2,$ while the family remain unchanged with probability $P_{u}=\left|r_0\right|^2+\left|r_3\right|^2=\cos\left(\theta\right)^2.$

Since, in the absence of Eve, the Alice's code never change the family of the Bell state, applying $S$ on travel qbit by Bob with $\theta\neq 0$ produces QBER until this QBER wiped with $S^\dagger$ by Alice. Any attempt of Eve, (measurements or operations) on travel qbit can disturb the procedure of producing and wiping QBER (hereafter we refer to this procedure as PW-QBER) and expose her.
So Eve must design an attack to reveal the codes without any direct measurement or operation on travel qbit.

 A promising class of attacks involves coupling an ancillary system to the traveling qubit during the forward transmission, entangling the two systems through a unitary operation, and subsequently attempting to extract information about the encoded bit from measurements on the ancilla during the backward transmission while introducing minimal disturbance to the protocol.

Since the action of a unitary operation on the joint system induces a generally non-unitary positive map on the traveling qubit, such attacks inevitably modify the PW-QBER and may therefore reveal Eve's presence. In the following, we show that the proposed protocol is secure against the most general form of ancilla-assisted attacks of this type.

We emphasize that the scrambling mechanism creates two independent ambiguities for Eve: one in the effective channel determined by Bob's scrambling operation and another in the meaning of Alice's encoding, which is randomized by her scrambling operation. Consequently, Eve is uncertain both about \textit{where} the information is encoded and \textit{what} the encoding represents. The interplay of these two uncertainties enhances the protocol's resilience against information-extraction attacks.

Before the security proof, let's discuss the form of $S$. From the first property, $S$ must producing QBER, so it should contain family-changing operators $X$ or $Y$. furthermore, Eve never knows about the mode of scrambling (DS,NS or MS modes), so $S$ must also contain the code operators $I$ or $Z$ to produces SA. The equal weighted of these two families of operators, scrambles the initial state into an equal superposition of the changed family (which produces QBER) and the unchanged family with the wrong code in some cases (SA will happens). Although the equal super position is not necessary, it makes the symmetry on QBER which prevent Eve to extract any information from the asymmetries of final results. Hereafter for simplicity we suppose $S=\frac{1}{\sqrt{2}}\left({X+Z}\right)=H$ as the Hadamard operator.
\subsection{General attack}   
Suppose Eve attaching two-dimensional ancilla system to traveling qbit and makes them entangled by general unitary operator $U_{et}$ where we have used subscribe $e$ and $t$ to refer $Eve$ and $Travel$ subsystem. She lets Alice to apply her operator $O_t$ which contain the codes and revert the evolution by $U^{\dagger}_{et}$ to be sure about the least disruption of travel qbit.
 
Generally, the local operator $O_t$ changes the entangled state of ancilla-travel and can be detected by local measurement on ancilla by Eve, But there are some important questions: is it possible for Eve to gain information without detection? what is the best strategy to gain more information with least detection?

Suppose Alice and Bob sharing $\ket{\beta_{pf}}_{ht}$ and without loss of generality, Eve uses $\ket{0}_e$ as an ancilla (Note that the general operator $U_{et}$ may include any local operator on ancilla qbit and it can transform $\ket{0}_e$ into any desired state). The final state after the attack is
\begin{equation}\label{psi_f}
	\ket{\psi}=\left(U^\dagger_{et}\otimes I_h\right)\left(I_e\otimes O_t \otimes I_h\right)\left(U_{et}\otimes I_h\right)\ket{0}_e \ket{\beta_{pf}}_{th}.
\end{equation}
So, the local state of Eve is
\begin{equation}\label{rho_e}
	\rho_e\left(O_t\right)=Tr_{th}\left(\ketbra{\psi}{\psi}{}\right)=\frac{1}{2}Tr_t\left(E\left(O_t\right)\varrho_{e} E^\dagger\left(O_t\right)\right)
\end{equation}
in which $E\left(O_t\right)=U^\dagger_{et} \left(I_e\otimes O_t \right)U_{et}$, $\varrho_{e}=\ketbra{0}{0}{e}\otimes I_t$ is initial state of ancilla and we have used $Tr_{h}\left(\ketbra{\beta_{pf}}{\beta_{pf}}{th}\right)=\frac{I_t}{2}$.

$O_t$ is the operator applied by Alice, which is either $H$ or $ZH$ when she decides to apply $H$ before code, and $I$ or $Z$ otherwise. The goal of Eve is to use the specific $U_{et}$ to gain more information about code. For this goal she must has optimal distinguishable states for different codes in all situations.

On of the good measures of distinguishability is the trace distance
\begin{equation}
	T\left(\rho_1,\rho_2\right)=\frac{1}{2}Tr\left(\sqrt{\left(\rho_1-\rho_2\right)^\dagger\left(\rho_1-\rho_2\right)}\right),
\end{equation}
which ranges from 0 for undistinguished states to 1 for perfect distinguishable.

By Bloch representation of density matrix 
$
	\rho=\frac{1}{2}\left(I+\vec{r}.\vec{\sigma}\right)
$
in terms of Block vector $\vec{r}$ and Pauli matrices $\vec{\sigma}$, it is easy to show that
$T\left(\rho_1,\rho_2\right)=\frac{1}{2}\left|\vec{r_1}-\vec{r_2}\right|$, in which $\vec{r_i}$ is Bloch vector of $\rho_i$.

We define 
\begin{align}\label{deff-Gamma-gamma}
	\Gamma_{i,j}&=\frac{1}{2}\left|\vec{r}\left(H^i\right)-\vec{r}\left(ZH^j \right)\right|\notag\\
	\gamma_{i}&=\frac{1}{2}\left|\vec{r}\left(Z^i\right)-\vec{r}\left(Z^iH \right)\right|,
\end{align}
in which $\vec{r}\left(O\right)$ is the Bloch vector of $\rho_e\left(O\right)$.

The quantities $\Gamma_{i,j}$ characterize Eve's ability to discriminate between the encoded bits and therefore serve as measures of her information gain. By contrast, $\gamma_0$ and $\gamma_1$ characterize the distinguishability of the same encoded bit across different scrambling modes. Since Eve does not know either the encoded bit or the active mode, a large $\gamma_i$ does not provide useful information; rather, it increases the variability of the quantum state associated with a given code and thus the ambiguity of its interpretation. Consequently, Eve seeks to maximize $\Gamma_{i,j}$ while simultaneously minimizing $\gamma_0$ and $\gamma_1$.

We have proven (see Appendix A) that, a meaningful attack on this protocol is impossible. Specifically, if an attack can perfectly reveal codes in the NS mode (e.g., achieving $\Gamma_{0,0} = 1$), it neither perfectly reveals codes in other modes ($\Gamma_{1,1} \leq 1$) nor achieves correctness in extracted codes more than half the time ($\Gamma_{0,1} \leq \frac{1}{2}$ and $\Gamma_{1,0} \leq \frac{1}{2}$). More precisely, when Eve attempts higher distinguishability in DS mode ($\Gamma_{1,1} \rightarrow 1$), she encounters increased ambiguity in interpreting results due to $\Gamma_{0,1} \rightarrow 0$ and $\Gamma_{1,0} \rightarrow 0$, which means code $I (Z)$ in NS mode becomes indistinguishable from code $Z(I)$ in DS or MS modes. Since scrambling is applied randomly, she cannot extract meaningful code information.
\section{Detection probability of Eve}
 In the absence of Eve, the prepared Bell state \(\ket{\beta_{pf}}\) is only modified by Alice's coding operations \(Z^a\), which preserve its family in both NS and DS modes. Thus, deviations in the state's family exceeding the system's normal threshold indicate Eve's presence. Using the shared state \(\ket{\beta_{pf}}\) and the final state \(\ket{\psi(a,m)}\) (from Eq.~\eqref{psi_f} with \(O_t = Z^a\) and mode \(m\)) that carries the code \(a\), the detection probability is given by
\begin{align}\label{Pd_c}
	Pd_c&=\sum_{m=\{NS,DS\}}p_m\sum_{i,a=0}^{1}p_a\left|\braket{\beta_{if\oplus 1}}{\psi(a,m)}{}\right|^2
\end{align}
in which $p_a$ is the probability of applying code $a$ and $p_m$ is the probability of running mode $m$ which is either $NS$ or $DS$ in \textit{coding mode}. We have another detection probability in this protocol, came from  the \textit{garbage} results in $MS$ mode which happens either when only Bob apply $H$ ($MS_B$) or only Alice apply $H$ ($MS_A$). We can write this detection probability as
\begin{align}\label{Pd_g}
	Pd_g&=\sum_{m=\{MS_A,MS_B\}}p_m\sum_{i,a=0}^{1}p_a\left|\braket{\beta_{p\oplus i \oplus a f\oplus i}}{\psi(a,m)}{}\right|^2.
\end{align}  
Note that, in the absence of Eve, \(MS_A \equiv MS_B\). Therefore, in the garbage cases, the effective operation reduces to \(Z^aH\), with
\(H=\frac{1}{\sqrt{2}}\sum_{i=0}^{1} Z^i X^{i\oplus 1}\).
This operation yields only the Bell states \(\ket{\beta_{p\oplus i\oplus a\,f\oplus i\oplus 1}}\), \(i=0,1\) (see Eq.~\eqref{local-pauli}). Consequently, the observation of any other Bell state indicates Eve's presence with probability \(Pd_{g}\).

 \section{Example: Double Control U attack}
In this section, we investigate a known effective attack against two-way quantum cryptography protocols: the DCNOT attack \cite{Lucamarini2014}. This attack is particularly effective against the message mode of the LM05 protocol, enabling Eve to perfectly reveal the correct secret codes. Eve executes the attack by coupling an ancilla qubit to the travel qubit. She applies CNOT gates both before and after Alice's station, consistently setting the travel qubit as the control and her ancilla qubit as the target.

In our protocol, since the coding operators are limited to $I$ and $Z$, they do not create superpositions on the travel qubit (Unlike $iY$ in LM05). Consequently, the original DCNOT attack is inapplicable: the CNOT gates cannot entangle the travel qubit with the ancilla under these conditions. Eve can overcome this limitation by changing the basis of code operators by a unitary transformation $U_t$. So she need  to apply $U_t$ immediately before the code operation and $U_t^\dagger$ after that.

Let us consider a generalized form of this attack. Eve first implements a controlled-\( U_e \) operation where the travel qubit serves as the control and the ancilla qubit as the target, conditionally applying \( U_e \) to the ancilla. She then applies a single-qubit transformation \( U_t \) to the travel qubit to change its basis right before code operation. After Alice applies her coding operator, Eve reverses this process by first applying the transformation \( U_t^\dagger \) operation, followed by controlled-\( U_e^\dagger \) on the travel qubit.
By expressing $U_e = U(\alpha,\beta,\theta)$ and $U_t = U(\alpha',\beta',\theta')$ in the general $SU(2)$ form defined in Eq.~\ref{SU2}, we have
\begin{align}
	U_{et}=\left(I_e\otimes U_t\right)\left(I_e\otimes\ketbra{0}{0}{t}+U_e\otimes\ketbra{1}{1}{t}\right).
\end{align}

\begin{figure}[!t]
	\includegraphics[width=0.8 \linewidth]{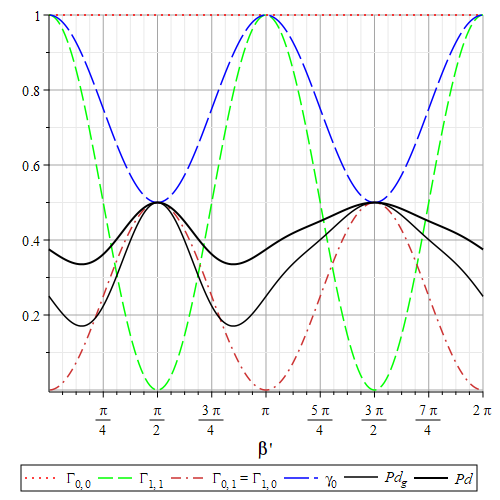}
	\caption{Dependence of the distinguishability measures $\Gamma_{0,0}$ (red dotted), $\Gamma_{1,1}$ (green dashed), $\Gamma_{1,0}=\Gamma_{0,1}$ (orange dash-dotted), and $\gamma_{0}$ (blue long-dashed), together with the detection probabilities $Pd_{g}$ (solid black) and $P_d=\frac{1}{2}\left(Pd_{c}+Pd_{g}\right)$ (solid thick black), on the parameter $\beta'$. The parameters are fixed at $\theta=\frac{\pi}{2}$, $\beta=\frac{\pi}{2}$, $\theta'=\frac{\pi}{4}$, and $\alpha'=\frac{\pi}{2}$.
	}\label{GammaijandPd}
\end{figure}

Using Eqs.~\eqref{psi_f}, \eqref{rho_e}, and \eqref{affin-map}, \(\Gamma_{0,0}\) in Eq.~\eqref{Gamma00} simplifies to
$
	\Gamma_{0,0}=\sin^{2}(2 \theta'){\left| \sin (\theta)\right|}\sqrt{1-\cos^{2}(\theta )\cos^{2}(\alpha)},
$
where only the parameter \(\theta'\) from \(U_t\) affects \(\Gamma_{0,0}\). Eve must therefore set \(\theta' = \pm\frac{\pi}{4}, \pm\frac{3\pi}{4}\) to ensure that the maximum distinguishability 1 is achievable. With this \(\theta'\) constraint and defining \(G(\alpha,\beta,\theta)={\left| \sin (\theta)\right|}\sqrt{1-\cos^{2}(\theta )\cos^{2}(\alpha)}\)
\begin{align}\notag\label{Gammaij}
	\Gamma_{0,0}&= G(\alpha,\beta,\theta)\\\notag
	\Gamma_{1,1}&=\sin^{2}(\phi')\, G(\alpha,\beta,\theta)\\
	\Gamma_{0,1}&=\Gamma_{1,0}=\frac{\cos^{2}(\phi')}{2}\,G(\alpha,\beta,\theta)\\\notag
	\gamma_{0}&=\gamma_{1}=\frac{1+\sin^{2}(\phi')}{2}\,G(\alpha,\beta,\theta)\notag
\end{align}
where \(\phi' = \alpha' - \beta'\). All \(\Gamma_{i,j}\) are proportional to \(G(\alpha,\beta,\theta)\), a function that attains its maximum value of 1 at \(\theta = \frac{\pi}{2}\). Interestingly, \(\theta = \frac{\pi}{2}\) corresponds to the family of off-diagonal unitaries \(U_e\) (termed the NOT family), which depend solely on \(\beta\) (for example $\beta=0$ and $\beta=\frac{\pi}{2}$ correspond to \(U_e=iY\) and \(U_e=iX\)  respectively). Eq.\ref{Gammaij} confirms the bounds established in Appendix~\ref{AppendixA}, namely,
$\Gamma_{1,0}, \Gamma_{0,1} \leq \frac{1}{2}$ and $\gamma_{0}, \gamma_{1} \geq \frac{1}{2}$.
It also makes explicit the fundamental trade-off faced by Eve: any attempt to increase the distinguishability quantified by $\Gamma_{1,1}$ necessarily leads to greater ambiguity in $\gamma_{0}$ and $\gamma_{1}$.
As an illustrative example, Fig. \ref{GammaijandPd} depicts an attack strategy for which perfect distinguishability is attainable. According to the analytical expressions derived above, achieving ($\Gamma_{1,1}=1$) requires the parameter choices ($\theta=\frac{\pi}{2}$) and ($\theta'=\frac{\pi}{4}$). Accordingly, Eve employs the NOT-family operation $U_e=iX\quad\left(\theta=\frac{\pi}{2},\beta=\frac{\pi}{2}\right)$,
together with the Hadamard-family operation
$U_t=\frac{i}{\sqrt{2}}\left(\sin(\beta')X+\cos(\beta')Y+Z\right)
\quad
\left(\theta'=\frac{\pi}{4},\alpha'=\frac{\pi}{2}\right).$
These parameter choices ensure that perfect code distinguishability is achievable. The remaining free parameter ($\beta'$) can then be varied to investigate the corresponding detection probabilities and the trade-off between Eve's information gain and the disturbance introduced into the protocol. Substituting these operators into Eqs.~\eqref{Pd_c} and \eqref{Pd_g} yields a detection probability of ($Pd_c=\frac{1}{2}$) and
\begin{align}
	Pd_g=\frac{1}{4}\left(\sin(\beta')^4+\sin(\beta')^3-\sin(\beta')\right).
\end{align} 
The minimum garbage detection probabilities, $Pd_{g}\simeq 0.17$ and whole detection probability $Pd\simeq 0.335$, occur at $\beta' \simeq 0.473$ and $\pi-0.473$. At these points, Eve can distinguish the code states perfectly in the NS mode ($\Gamma_{0,0}=1$) and with success probability $\frac{1}{2}(1+\Gamma_{1,1})\simeq 0.89$ in the DS mode. However, she is nearly unable to distinguish code $I$ ($Z$) in the NS mode from code $Z$ ($I$) in the DS mode, as indicated by $\Gamma_{0,1}=\Gamma_{1,0}\simeq 0.1$. This ambiguity is reflected in the large values $\gamma_{0}=\gamma_{1}\simeq 0.89$.  Consequently, even under the optimal strategy, Eve's probability of correctly identifying the encoded bit is only $\frac{1}{2}(1+\Gamma_{0,1})\simeq 0.55$, which is merely $5\%$ above random guessing, while exposing her presence with probability 0.33. Moreover, no choice of attack parameters allows Eve to evade detection completely, as $Pd_g$ and $Pd_c$, remain strictly nonzero (see Fig.~\ref{GammaijandPd}).

\section{Conclusion}

In this work, we have introduced a novel two-way quantum key distribution protocol based on a scrambling-operator mechanism. The protocol possesses several distinctive features. First, the secret key is not encoded directly in the traveling qubit but rather in the nonlocal correlations shared between Alice and Bob. This significantly restricts attacks that attempt to extract information from the transmitted quantum state alone.

Second, unlike some existing QKD protocols, the control mode remains completely private and is never revealed, either during the protocol execution or in the final sifting stage for "code" rounds. Consequently, adaptive attacks and strategies designed to exploit or remain hidden within the control mode are ineffective against the proposed protocol.

Third, a subset of the raw key is independently identified by Bob as ``garbage'' rounds and can be used for eavesdropper detection. This classification is performed solely by Bob at the end of the protocol and requires only Alice's disclosure of her scrambling operation. No additional information exchange is needed to identify the code and garbage rounds. Moreover, the rounds labeled as ``code'' can also be utilized for eavesdropper detection. As a result, the protocol enables security verification without sacrificing key bits, mode-dependent measurements, or additional communication between the legitimate parties.

Finally, we have investigated a general class of ancilla-assisted attacks, in which Eve couples an ancillary system to the traveling qubit and attempts to infer the encoded information from subsequent measurements. We have shown that any meaningful information gain is necessarily accompanied by detectable disturbances, demonstrating the robustness of the proposed protocol against this broad class of attacks.

\textbf{Author Contributions:} M. Annabestani conceived the protocol, developed the theoretical framework, established the security proof, and wrote the manuscript. S. Sarshar contributed to the preparation of the manuscript and discussions of the results.

 \section{Appendix. A}\label{AppendixA}
 The Bloch vector of $\rho_e\left(O_t\right)$ in Eq. \ref{rho_e} can represent as 
 \begin{equation}\label{affin-map}
 	\vec{r}\left(O_t\right)=M\left(O_t\right)\vec{r_0}
 \end{equation}
 where $M\left(O_t\right)=P\left(O_t,O_t\right)$ is  3 dimensional matrix  and $\vec{r_0}$ is Bloch vector of ancilla $\varrho_{e}$. In fact $M\left(O_t\right)$ is the special form of $P\left(O_1,O_2\right)$ which it's elements are
 \begin{equation}
 	P_{\alpha,\beta}\left(O_1,O_2\right)=\frac{1}{4}Tr_{et}\left[\left(\sigma_\alpha\otimes I_t\right)E\left(O_1\right)\left(\sigma_\beta\otimes I_t\right)E(O_2^\dagger)\right],
 \end{equation}
 
 Suppose Eve design an attack (set $U_{et}$) to reveals the codes perfectly in NS mode (Alice and Bob do not apply scramble) which means
 \begin{equation}\label{Gamma00}
 	\Gamma_{0,0}=\frac{1}{2}\left|\vec{r}\left(I\right)-\vec{r}\left(Z\right)\right|=\frac{1}{2}\left|\vec{r_0}-M\left(Z\right)\vec{r_0}\right|=1.
 \end{equation} 

 Since she doesn't have any ideal about the modes are running, her attack must be able to distinguished codes in all modes which means $\Gamma_{0,1}, \Gamma_{1,0}$ and $\Gamma_{1,1}$ should be maximize. Lets see $\Gamma_{0,1}$ in details
 
 \begin{equation}\label{Gamma01}
 	\Gamma_{0,1}=\frac{1}{2}\left|\vec{r}\left(I\right)-\vec{r}\left(ZH\right)\right|=\frac{1}{2}\left|\vec{r_0}-M\left(ZH\right)\vec{r_0}\right|
 \end{equation} 
in which we have used $M\left(I\right)=I$. 

By use of $ZH=\frac{1}{\sqrt{2}}\left(iY+I\right)$ and some simplification
\begin{align}\label{MHZ}
	M\left(ZH\right)=\frac{1}{2}\big(I+M\left(Y\right)+P\left(Y,I\right)+P\left(I,Y\right)\big)
\end{align} 
So from Eq. \eqref{Gamma01}
\begin{equation}\label{Gamma01finall} 
	\Gamma_{0,1} = \frac{1}{4} \left|\vec{r_0} - \big(M\left(Y\right)+P\left(Y,I\right)+P\left(I,Y\right)\big)\vec{r_0}\right|. 
\end{equation} 
It is clear from Eq. \eqref{affin-map} that $\left|M\left(ZH\right)\vec{r_0}\right| \leq 1$. Therefore, by applying the triangle inequality ($|\vec{a}+\vec{b}| \leq |\vec{a}| + |\vec{b}|$) with $\vec{a} = \vec{r_0}$ and $\vec{b} = \big(M\left(Y\right)+P\left(Y,I\right)+P\left(I,Y\right)\big)\vec{r_0}$ in Eq.\ref{MHZ}, we find that
$\left|\big(M\left(Y\right)+P\left(Y,I\right)+P\left(I,Y\right)\big)\vec{r_0}\right| \leq 1$.
By this fact and using triangle inequality again, Eq. \eqref{Gamma01finall} implies 
\begin{align}
	\Gamma_{0,1}\leq\frac{1}{2}.
\end{align}
This means that in at least half of the cases, the code Eve obtains by interpreting her results is incorrect. In other words, Eve has no knowledge of the outcome and is entirely uncertain whether the state corresponds to code I in NS mode or to code Z in DS or MS mode.

Furthermore, by using inverse triangular inequality ($| |\vec{a}| - |\vec{b}|| \leq|\vec{a}-\vec{b}|$) with $\vec{a}=\vec{r}\left(I\right)-\vec{r}\left(Z\right)$ and $\vec{b}=\vec{r}\left(I\right)-\vec{r}\left(ZH\right)$
\begin{align}
	|\Gamma_{0,0}-\Gamma_{0,1}|\leq \gamma_1
\end{align}
which implies $\gamma_1\geq \frac{1}{2}$. This indicates that when obtaining code Z ( $\gamma_1$ in Eq. \eqref{deff-Gamma-gamma}), Eve faces ambiguity. Specifically, in at least half of the cases, her attack produces inconsistent results for the same code (Z), preventing her from validating the obtained code. 

In the same way, for
\begin{equation}\label{Gamma10}
	\Gamma_{1,0}=\frac{1}{2}\left|M\left(H\right)\vec{r_0}-M\left(Z\right)\vec{r_0}\right|,
\end{equation}
we have  
\begin{align}\label{MH}
	M\left(H\right)=\frac{1}{2}\big(M\left(X\right)+M\left(Z\right)+P\left(X,Z\right)+P\left(Z,X\right)\big).
\end{align}
Therefore, by the constrain of $|M\left(H\right)\vec{r_0}|\leq 1$ and the perfect distinguishable code, assumed in Eq. \eqref{Gamma00} ( $|M\left(Z\right)\vec{r_0}|= 1$), we have
\begin{align}\label{MH_ineq}
	\left|\big(M\left(X\right)+P\left(X,Z\right)+P\left(Z,X\right)\big)\vec{r_0}\right| \leq 1.
\end{align}
So by substitution of Eq. \eqref{MH} into Eq. \eqref{Gamma10}
\begin{equation}\label{Gamma10finall}
	\Gamma_{1,0}=\frac{1}{4}\left|\big(M\left(X\right)+P\left(X,Z\right)+P\left(Z,X\right)\big)\vec{r_0}-M\left(Z\right)\vec{r_0}\right|.
\end{equation}
which by triangular inequality with $\vec{a}=\big(M\left(X\right)+P\left(X,Z\right)+P\left(Z,X\right)\big)\vec{r_0}$ and $\vec{b}=M\left(Z\right)\vec{r_0}$ and Eq. \eqref{MH_ineq} has upper bound as 
\begin{align}
	\Gamma_{1,0}\leq\frac{1}{2}.
\end{align}

%

Thus, the maximum value of both $\Gamma_{0,1}$ and $\Gamma_{1,0}$ is at most $\frac{1}{2}$. This maximum is achieved when, $\vec{r}_{IY} = \left(M(Y) + P(Y,I) + P(I,Y)\right)\vec{r_0}$ (see Eq.~\eqref{Gamma01finall}) is antiparallel to $\vec{r_0}$, and $\vec{r}_{XZ} = \left(M(X) + P(X,Z) + P(Z,X)\right)\vec{r_0}$ (see Eq.~\eqref{Gamma10finall}) is parallel to $\vec{r_0}$ (noting that Eq.~\eqref{Gamma00} assumes $M(Z)\vec{r_0}$ is antiparallel to $\vec{r_0}$). These conditions cause $M(H)$ (Eq.~\eqref{MH}) and $M(ZH)$ (Eq.~\eqref{MHZ}) to vanish, yielding:
\begin{equation}\label{Gamma11}
		\Gamma_{1,1} = \frac{1}{2} \left| M(H) \vec{r_0} - M(ZH) \vec{r_0} \right| = 0.
\end{equation}
Conversely, if $\vec{r}_{IY}$ is parallel and $\vec{r}_{XZ}$ antiparallel to $\vec{r_0}$, then $\Gamma_{1,1} = 1$ while $\Gamma_{0,1} = \Gamma_{1,0} = 0$.

Consequently, any attempt by Eve to increase $\Gamma_{1,1}$ necessarily reduces $\Gamma_{0,1}$ and $\Gamma_{1,0}$ below their marginal value of $\frac{1}{2}$, and vice versa. This trade-off cannot be circumvented by adopting a mode-dependent attack strategy, since Eve does not know whether a given round belongs to the code or garbage class at the time of her interaction with the quantum system. In particular, she has no prior knowledge of Bob's application of $S$ and determining Alice's application of $S^{\dagger}$ is fundamentally imperfect. The operator $S^{\dagger}$ is deliberately chosen to exhibit both scrambling ability and a nonzero QBER, so that the resulting quantum states possess nonzero overlap with both the correct and incorrect code hypotheses. Therefore, distinguishing the application of $S^{\dagger}$ cannot be achieved with certainty, even when Eve is granted arbitrary quantum resources, including higher-dimensional ancillas, collective measurements, or multipartite entanglement. As a result, any gain in distinguishability for one mode is necessarily accompanied by a loss in distinguishability for the others.

\bibliography{QKDreferences.bib}

\end{document}